\documentclass[apj]{emulateapj}
\usepackage{amsmath}
\usepackage{graphicx}
\usepackage{amssymb}
\usepackage{times}

\begin{document}

\title{Conversion Measure of Faraday Rotation-Conversion with application to Fast Radio Bursts}

\author{Andrei Gruzinov}
\affiliation{Physics Dept., New York University,  726 Broadway, New York, NY 10003, USA}

\author{Yuri Levin}
\affiliation{Center for Theoretical Physics, Department of Physics,
Columbia University, New York, NY 10027, USA}
\affiliation{Center for Computational Astrophysics,
Flatiron Institute, New York, NY 10010, USA}
\affiliation{School of Physics and Astronomy,
Monash University, VIC 3800, Australia}

\begin{abstract}

Faraday Rotation-Conversion is the simultaneous rotation of all three Stokes polarization parameters $Q$, $U$, $V$ as an electromagnetic wave propagates through a magnetized plasma. In this regime the Faraday plasma screen is characterized by more than just a Rotation Measure. We define the Conversion Measure that characterizes the wavelength-dependent conversion between the linear and circular polarization. In a cold plasma, the conversion occurs at the localized regions along the wave's path, where the large-scale magnetic field is perpendicular to the propagation direction. We show that the number of these regions along the line
of sight through the screen, and their individual contributions to the conversion measure, can be inferred from the polarization measurements. We argue that the simultaneous measurement of wavelength-dependent linear and circular polarization might give an important insight into the magnetic-field geometry of the Faraday screen in FRB121102 and other repeating
Fast Radio Bursts.

\end{abstract}
\keywords{polarization}

\maketitle

\section{Introduction}
\label{sec:intro}
Faraday conversion between the linear and circular polarizations is thought to be responsible for producing the measured circular polarization of some radio sources in galactic nuclei [e.g.,~\cite{RB},~\cite{HL},~\cite{Bower},~\cite{JO}].
More recently, Vedantham \& Ravi (2018) have pointed out that Faraday Conversion (FC)
%between linear and circular polarizations
might be relevant for Fast Radio Bursts (FRB), especially for the repeating FRB121102. This source, discovered by Spitler et al.~(2014, 2016) and studied extensively since then, provides an important test bed for the models of the origin of Fast Radio Bursts [e.g., Beloborodov 2017, Waxman 2017, Thompson 2017, Margalit \& Metzger 2018]. The FC reflects the birefringence of the plasma surrounding the FRB, and is potentially a sensitive probe of
the magneto-ionic environment of this enigmatic source, since (i) the measured linear polarization is close to 100\%, with extremely high Rotation Measure of $1.46\times 10^5\hbox{rad}/\hbox{m}^2$ (Michilli et al.~2015), (ii) while
the currently measured circular polarization is consistent with zero at frequencies higher than $~5$GHz, this may not be so at lower frequencies and is potentially measurable.

Vedantham \& Ravi (2018) stated that in the cold plasma the FC is suppressed, because the normal-mode waves are nearly circular, and that
instead the conversion would happen in a relativistic plasma. 
However, in this paper we show that FC can occur efficiently in cold plasma
near the points where the Faraday Rotation frequency is close to zero, i.e.~near the "field reversals" where the magnetic field component along 
the line of sight is zero. The importance of reversals in Faraday conversion has been studied in the past starting from Cohen (1960) and Zheleznyakov \& Zlotnik (1964). Ruzskowski \& Begelman (2002) explored the role of reversals in the context of circular polarization from galactic nuclei. Melrose et al.~(1995), Broderick \& Blandford (2010), and Melrose (2010) give a comprehensive discussion of passage through a reversal in terms of rotation of the Poincare sphere. 

In this paper we identify several novel diagnostics for polarization state of the pulse traveling through a series of reversals. 
We present a simple theory of the effect in \S~\ref{sec:theory}, and we use the theory in \S~\ref{sec:appl} to predict what might be seen in FRB 121102 if the polarization is measured at sufficiently low frequencies $\nu$. 

In this work we focus on cold plasma. Our theoretical results can be summarized as follows. A 100\% linear polarized wave, after passing through several field reversals in a magnetized plasma (called Faraday screen in this context) becomes partially circularly polarized. 
%The conversion occurs near the field reversals, i.e.~near the points where the magnetic field is perpendicular to the line of sight. 
After a 
passage through the reversal, the circular polarization $V$ oscillates quasiperiodically as a function of $\lambda ^2$, where $\lambda=c/\nu$ is the wavelength. For the screen parameters and wave frequencies which are expected to be relevant for FRB 121102:
\begin{itemize}

\item The rms value of $V$ oscillations is given by 
\begin{equation}
\langle \Pi _0\rangle={\rm CM}~\lambda^2,
\end{equation}
where $\Pi _0\equiv |V|/{I}$ is the degree of circular polarization. Total intensity $I=1$ is assumed throughout the paper, and we have defined a Conversion Measure CM measured in units of $1/{\rm m}^2$. The result is valid over a finite wavelength interval, where conversion is small and rotation is large, 
\begin{equation}\label{apco}
{\rm CM}~\lambda^2\ll 1,~~~{\rm RM}~\lambda^2\gg 1.
\end{equation}
Here RM is the usual Faraday rotation measure. We also show that for a passage through a single smooth reversal, an asymptotically exact expression
is
\begin{equation}\label{exco}
   \langle \Pi _0\rangle=\sqrt{2\left(e^{-c^2/2}-e^{-c^2}\right)},~~~c\equiv {\rm CM}~\lambda^2. 
\end{equation}
This equation is valid with a high degree of precision for ${\rm RM}~\lambda^2\gg 1$ and arbitrary ${\rm CM}~\lambda^2$.
{We note that this solution was already found in Zheleznyakov \& Zlotnik (1964) in a slightly different form\footnote{We thank Harish Vedantham for poining this out after the first version of this paper appeared on the arxiv.  While the $\lambda$-dependence for the conversion angle follows directly from Zheleznyakov and Zlotnik's results, it was not emphasized in that paper and Conversion Measure is defined here for the first time. }.}
\item For FRB 121102 the expected conversion measure is 

\begin{equation}
{\rm CM}\sim 1{\rm m}^{-2}
\end{equation}
giving the rms degree of circular polarization
\begin{equation}
\langle\Pi _0\rangle\sim 10\%
\end{equation}
at $\nu \sim 1$GHz, but with very high uncertainty, as explained in \S~\ref{sec:appl}.

\item The number of different qusiperiods, i.e.~the number of peaks of the Fourier transform of $V$ as a function of $\lambda ^2$ is equal to the number of reversals in the large-scale magnetic field. The locations of the peaks represent the Rotation Measures of the reversal points. This feature survives if the field has small-scale fluctuations due to a turbulent cascade at short wavelengths. Even though in this case there is a multitude of reversal points, they are strongly clustered around the 
reversals of the large-scale field and each of the clusters produces a potentially measurable quasiperiod.

\end{itemize}

\section{Faraday Rotation-Conversion}
\label{sec:theory}

Propagation of an electromagnetic wave in a Faraday screen changes all three Stokes polarization parameters $Q$, $U$, $V$. We will assume that the wave propagates along $z$ and  take $Q^2+U^2+V^2=1$. Then $Q=\pm 1$ is 100\% linear polarization along $x,y$; $U=\pm 1$ is 100\% linear along $x\pm y$; $V=\pm 1$ is 100\% right/left circular polarization. 

Assuming the plasma is cold, 
%for simplicity (when observations call, this assumption can be dropped), 
we have a simple equation for the polarization evolution in the screen [eg., Sazonov 1969, an ab initio derivation is given in the Appendix]:
\begin{equation}\label{tra}
{\bf P}'=\boldsymbol\Omega \times {\bf P}, ~~~~~{\bf P}\equiv (Q,U,V).
\end{equation}
The polarization vector ${\bf P}$ rotates with angular velocity $\boldsymbol\Omega$ {(measured in units of $\hbox{m}^{-1}$)} as the wave propagates along $z$, and $'\equiv d/{dz}$. The components of the polarization rotation-conversion rate 
\begin{equation}
\boldsymbol\Omega \equiv (g,h,f)
\end{equation}
are the Faraday rotation rate 
\begin{equation}
f=-\frac{1}{c}\frac{\omega_p^2\omega_B}{\omega^2}\hat{B}_z
\end{equation}
and the Faraday conversion rate
\begin{equation}
h+ig=-\frac{1}{2c}\frac{\omega_p^2\omega_B^2}{\omega^3}(\hat{B}_x+i\hat{B}_y)^2,
\end{equation}
where $\hat{B}\equiv {\bf B}/{B}$ is the unit vector along the magnetic field of the screen ${\bf B}$, $\omega$ is the angular frequency of
the wave, and 
\begin{equation}
\omega_p^2=\frac{4\pi ne^2}{m}, ~~~~~\omega_B=\frac{eB}{mc}
\end{equation}
are the plasma and Larmor frequencies in the screen.

It is important to note the hierarchy of the components of the angular velocity $\boldsymbol\Omega$ responsible for rotation-conversion:
\begin{equation}
\frac{g}{f}\sim\frac{h}{f}\sim \frac{\omega_B}{\omega}=\frac{\nu_B}{\nu}=\frac{2.8B_G~{\rm MHz}}{\nu},
\end{equation}
where $B_G\equiv {B}/{1 {\rm G}}$. {It is important to note, however, that these estimates are only valid if $B_x\sim B_y\sim B_z$, and may not hold in all regions along the path.} %and we use the same convention in what follows. 
If, say, $B\sim 1$ mG and $\nu \sim 1$ GHz, we have ${g}/{f}\sim 3\times 10^{-6}$ and the angular velocity ${\boldsymbol\Omega}$ is nearly aligned with the $V$-axis. 

Suppose we are interested in the evolution of an initially linearly polarized pulse, as is the case with the repeating radio bursts from FRB121102. The polarisation vector $\boldsymbol P$ is initially in the $U-Q$ plane and therefore we may consider the rotation of the whole plane with $\boldsymbol{\Omega}$ as the angular velocity. Clearly, this is equivalent to considering rotation of the vector normal to the $U-Q$ plane, i.e.~rotation of the unit vector $\hat{V}$ that represents initially purely circular polarization. We can immediately see that as 
$\hat{V}$ and the $U-Q$ plane rotate around the $\boldsymbol{\Omega}$, they turn at the most by the angle $\theta_{\rm max}\simeq 2\sqrt{g^2+h^2}/f\ll 1$, and thus the 
circular polarization $V$ of any initially linearly polarized pulse does not exceed $\theta_{\rm max}$.
%\begin{equation}
%\boldsymbol\Omega \approx (0,0,f)
%\end{equation}
%only rotates the linear polarizations $Q$, $U$. 
This is the Faraday rotation regime -- the common case in astrophysics, with negligible conversion between linear $(Q,U)$ and circular $V$ polarizations. In this regime the Faraday screen is fully characterized by a single parameter -- the rotation measure
\begin{equation}
{\rm RM}\equiv \frac{1}{2\lambda^2}\int dz~f=8.1\times 10^5 \frac{\rm rad}{{\rm m}^2}\int {dz\over{\rm pc}}{n\over {\rm cm}^{-3}}{B_z\over{\rm G}},
\end{equation}
relating initial and final linear polarizations:
\begin{equation}
(Q+iU)|_f=e^{2i{\rm RM}~\lambda^2}(Q+iU)|_i.
\end{equation}

\subsection{Adiabatic Invariant}
\label{sec:adi}
It is clear that FC will remain small, so long as the angle between the angular velocity vector $\boldsymbol{\Omega}$ and the $V$-axis  remains small. 
This statement can be made with greater rigor by noting that the polarization transfer equation (\ref{tra}) has an adiabatic invariant
\begin{equation}
P_\parallel\equiv \hat{\Omega}\cdot {\bf P}={\rm inv},~~~~~{\rm RM}~\lambda^2\gg 1,
\end{equation}
because
\begin{equation}
P_\parallel'=\hat{\Omega}'\cdot {\bf P}\rightarrow\hat{\Omega}'\cdot \langle{\bf P}\rangle=\hat{\Omega}'\cdot (P_\parallel \hat{\Omega})=0.
\end{equation}
This invariant has been expensively discussed in the literature (e.g., Melrose 2010). One can show that the invariant is conserved so long as 
\begin{equation}
    \left|\hat{\boldsymbol\Omega}'\right|\ll \Omega,
\end{equation}
i.e. when the Faraday rotation is faster than the rotation of direction of $\boldsymbol\Omega$. In that case, the vectors that were initially in the $U-Q$ plane, remain nearly perpendicular to $\boldsymbol\Omega$ and, if the latter remains close to the $V$-axis, the circular polarization remains very small.
%No matter how thick the screen, if the magnetic field in the screen changes slowly, as compared to the Faraday rotation rate, a linear-polarized wave stays linear-polarized.

%It is clear that FC will remain small, so long as the angle between the angular velocity vector $\boldsymbol{\Omega}$ and the $V$-axis  remains small. 
At first glance, for $\nu_B/\nu\ll 1$ the $V-\boldsymbol\Omega$ alignment and adiabaticity are always satisfied in cold plasma. However, this argument is flawed. As the pulse travels along the line of sight, it is likely to encounter field reversals, i.e.~the locations where the field is perpendicular to the line of sight, and both $B_z$ and $f$ are zero. At or near these locations,  $\boldsymbol\Omega$ is
strongly misaligned with the $V$-axis and its direction changes rapidly
as the pulse travels through the reversal. Therefore, the adiabaticity
can be broken, in which case the pulse can develop a substantial circular polarization. In the next section we discuss in more detail
the FC as the pulse crosses the field reversal.

%We show in \S\S \ref{sec:con}, \ref{sec:tur}  that at high rotation measure, ${\rm RM}~\lambda^2\gg 1$, but still small $\frac{\nu_B}{\nu}\ll 1$, a significant conversion between linear $(Q,U)$ and circular $V$ polarizations can occur provided  
%\begin{equation}\label{sco}
%({\rm RM}~\lambda^2)^{1/2}~\frac{\nu_B}{\nu}\gtrsim 0.1,
%\end{equation}
%and this condition might be met by FRB121102 at $\nu \lesssim 1$ GHz.

\subsection{FC at field reversals. Conversion measure.}
\label{sec:con}

Consider now the passage of a linearly polarized pulse near a field reversal at $z=0$.
It is convenient to choose the $x$-axis in the direction of the magnetic field at the reversal. Near the reversal, the angular frequency of the $\boldsymbol P$-rotation is given by 
\begin{equation}\label{linear}
    \boldsymbol\Omega=(0,h,f'z),
\end{equation}
where we consider $f'$, $h$ as constants. Furthermore, we assume that
this approximation is valid for a range of $-z_0<z<z_0$ such that $f'z_0\gg h$, i.e.~we assume that the pulse is in the Faraday-rotation limit as it both enters and exits the reversal. In this approximation, the evolution of the polarization vector is entirely characterized by a single dimensionless parameter, 
\begin{equation}
\xi=h^2/f'.
\label{keypar}
\end{equation}

In Figure 1, we show several examples of evolution of the conversion angle $\theta(z)$
as the pulse passes through the reversal. Here $\theta$ the angle by which the plane  of linearly-polarized Stokes vectors $Q-U$ (or, equivalently, a vector perpendicular to this plane) rotates as a result of FC. The circular polarization $|V|<\sin{\theta}$. The figure also shows the evolution of the angle $\alpha(z)$ by which the angular velocity $\boldsymbol\Omega$ turns during the passage. It is clear that for $\xi\gg 1$, the orientation of the plane follows adiabatically that of $\boldsymbol\Omega$. The circular polarization achieves its maximum at $z=0$, but then the plane flips into alignment with the original $Q-U$ plane and the circular polarization becomes small again. For $\xi\ll 1$, the rate of Faraday Rotation passes zero so quickly that the FC does not have time to occur [see e.g., Melrose et al.~(2010) or Broderick \& Blandford (2010)]. 

\begin{figure}
 \includegraphics[width=0.5\textwidth]{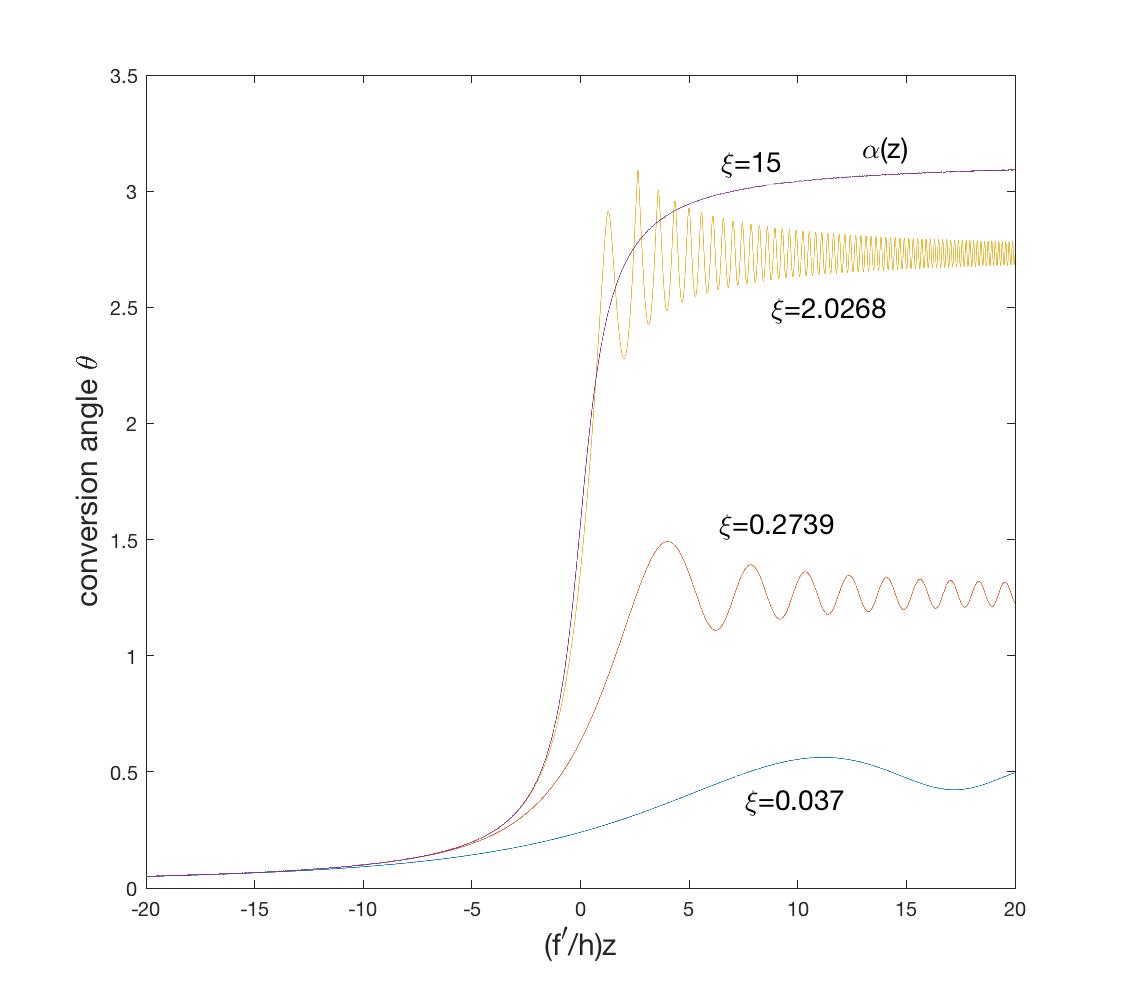}
\caption{Evolution of the conversion angle during a passage through a reversal. Plotted on the horizontal is the coordinate $z$ in terms of the characteristic length $h/f'$. With increasing $\xi$ the passage through a reversal becomes more gradual, reaching the adiabatic regime where the conversion angle tracks the angle by which $\hat{\boldsymbol\Omega}$ rotates. {The $\xi=15$ curve is indistinguishable from $\alpha(z)$, the angle by which the angular velocity $\boldsymbol\Omega$ turns during the passage}. The conversion is large for intermediate values $\xi\sim 1$. The curves are qualitatively similar to those in Figs.~(2)--(4) of Melrose et al.~(1995).} 
\label{fig:dynamics}
\end{figure}

It is possible to find the final value of the conversion angle $\theta_f=\theta(z=+\infty)$ analytically, by solving Eq.~(\ref{tra}) using a Laplace transform; {\rm it was also obtained by Zheleznyakov \& Zlotnik (1964) using a different method}. The answer  is given by
\begin{equation}
    \theta_f(\xi)=\arccos\left(2e^{-\pi\xi/2}-1\right).
\end{equation}
The greatest Faraday conversion occurs at values $\xi\sim 1$, with the angle $\theta_f$ reaching $\pi/2$ at $\xi=(2/\pi)\log 2$. At this value of $\xi$ a full conversion between the linear and circular polarizations is possible.
The RMS amplitude of $V$ is given by
\begin{equation}
    \langle\Pi_0\rangle ={1\over \sqrt{2}}\sin\theta_f=\sqrt{2\left(e^{-\pi\xi/2}-e^{-\pi\xi}\right)}.
\end{equation}
 
For small values of $\xi$, 
\begin{equation}
    \theta_f\sim \sqrt{2\pi\xi}
    \label{scaling1}
\end{equation}
and this turns out to be a good approximation for $\theta_f<1$rad. It is easy to understand qualitatively where this scaling is coming from. Near the reversal, $\boldsymbol P$ makes one Faraday rotation over the lengthscale $\Delta z\sim 1/\sqrt{f'}$. During this interval, the conversion angle is $\theta\sim h\Delta z\sim \sqrt{\xi}$. 

%The polarization transfer equation (\ref{tra}) with $\boldsymbol\Omega$ from (\ref{linear}) is of Laplace type -- its solution can be represented by contour integrals. The contour integrals can be calculated at $z=\pm \infty$, giving the exact expression (\ref{exco}). 

%As was explained in the previous section, the only way to have large conversion is to break the adiabatic invariant. This happens when the rate of Faraday rotation becomes small, namely near the zeros of $B_z$. Such zeros are indeed expected if the Faraday screen is created by intrinsic magnetic field of the plasma, rather then by the magnetic field of a star.

%One also needs a small rate of conversion. Since simultaneous zeros of $B_x$, $B_y$, $B_z$ are unlikely, at small frequencies, when the conversion rate becomes large, the adiabatic invariance must be restored. This low-frequency regime will be described separately in \S~\ref{sec:lcm}.

Assume now, for simplicity, that the coherence length of the magnetic field is comparable to the screen thickness $l$. If $B_z(z=0)=0$, then
\begin{equation}
f(z)\sim f_0\frac{z}{l}, ~~~~~|z|\lesssim l,~~~~~f_0\sim \frac{{\rm RM}~\lambda^2}{l},
\end{equation}
where $f_0$ is the characteristic value of $f$ far from the reversal.
%One Faraday rotation occurs over the distance 
%\begin{equation}
%z\sim \left(\frac{l}{f}\right)^{1/2},
%\end{equation}
From Eq.~(\ref{scaling1}) we see that the produced circular polarization (starting from 100\% linear) is about
\begin{equation}
V\sim \frac{g}{f_0}(lf_0)^{1/2}\sim \frac{\nu_B}{\nu}({\rm RM}~\lambda^2)^{1/2}.
\end{equation}

\begin{figure}
 \includegraphics[width=0.5\textwidth]{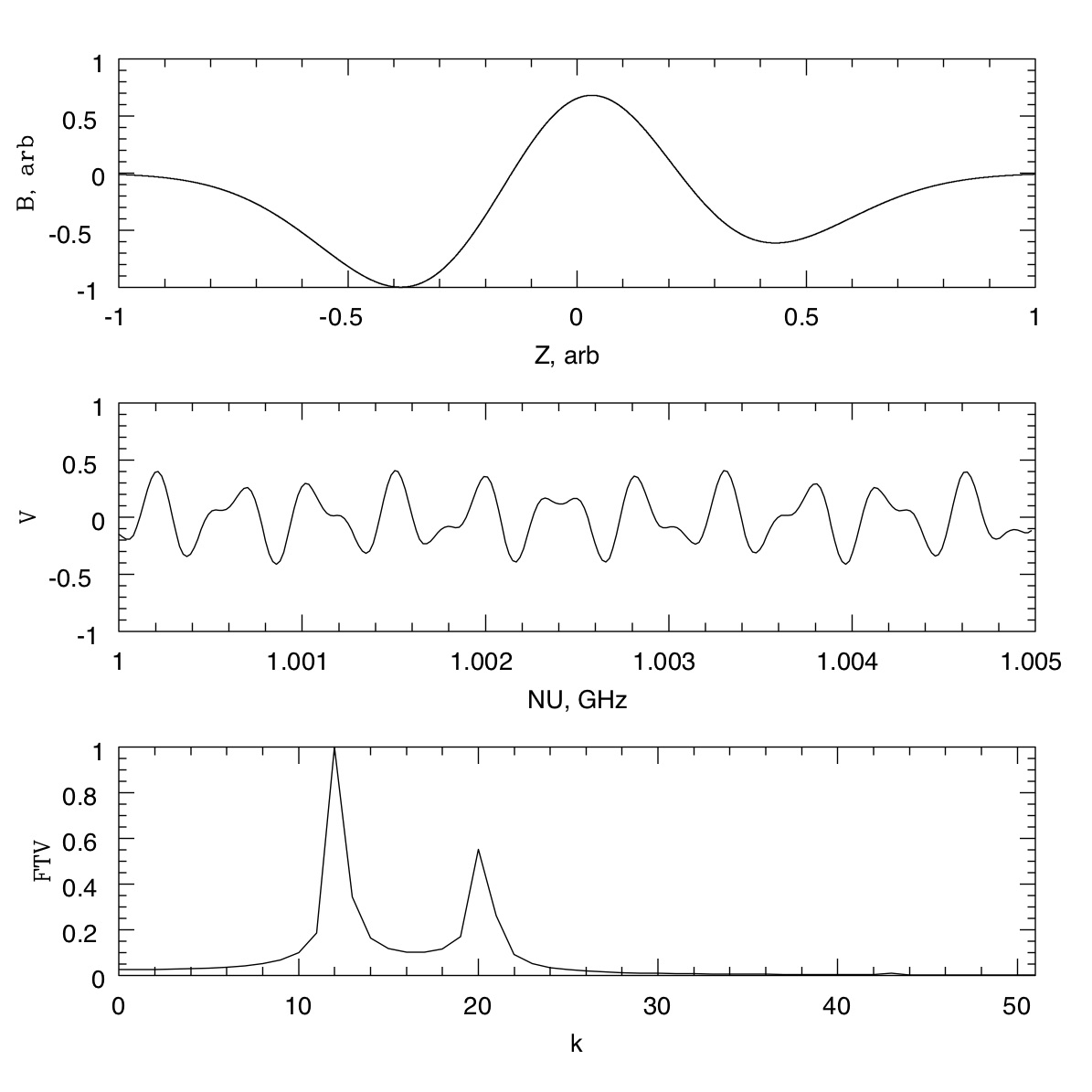}
\caption{Upper panel: parallel magnetic field $B_z$ vs $z$ in the Faraday sceen. Middle panel: circular Stokes parameter $V$ vs frequency $\nu$ after passage through the screen. Lower panel: Fourier transform of $V$ with respect to $\lambda^2$. The locations of the two peaks 
correspond to the Rotations Measures of the two field reversals. }
\label{fig:smoo}
\end{figure}

For small conversion angles, the rms value of $V$ is given by
\begin{equation}
    \langle \Pi_0\rangle={\rm CM}~\lambda^2,
    \label{CM2}
\end{equation}
while for large conversion angles,
\begin{equation}\label{exco1}
   \langle \Pi _0\rangle=\sqrt{2\left[e^{-\left({\rm CM} ~\lambda^2\right)^2/2}-e^{-\left({\rm CM} ~\lambda^2\right)^2}\right]}.
\end{equation}

Equation (\ref{CM2}) defines the conversion measure, 
%for a single reversal, 
which in the current simple example is related to the rotation measure ${\rm RM}\sim lf_0/\lambda^2$ as follows:
\begin{equation}\label{cme}
{\rm CM}\sim \frac{\nu _B}{c}~{\rm RM}^{1/2}\sim 10^{-2}\frac{1}{{\rm m}^2}~{\rm RM}^{1/2}_{\rm m}~B_{\rm G}.
\end{equation}

We have confirmed Eq.~(\ref{CM2}) by many dozens of numerical integrations of the polarization transfer equation (\ref{tra}), using overall screen parameters of \S~\ref{sec:appl} expected in FRB121102, but with different magnetic fields. So long as there are zeros of $B_z$, and so long as the applicability condition (\ref{apco}) is satisfied, Eq.~(\ref{CM2}) works. But we must still clarify what exactly is the rms circular polarization $\langle\Pi_0\rangle$. 

To this end consider again a passage of the linearly polarized pulse through a screen with a reversal, entering the screen at $z=-a$.
The Faraday rotation angle is given by 
%screen with a single zero of $B_z$: $n=$const, $B_x=$const, $B_y=0$, $B_z\propto z$. Then 
\begin{equation}
\int_{-a}^z dz_1~f(z_1)={f_0\over 2l}\left(z^2-a^2\right).
%~~~~~\alpha={\rm const}.
\end{equation}
Neglecting conversion, Eq.(\ref{tra}) gives 
\begin{equation}
Q+iU=e^{i(f_0/2l)(z^2-a^2)},
\end{equation}
%assuming that a 100\% linear polarization along $x$ enters the screen from below at $z=-a$. 
From Eq.(\ref{tra}), the circular polarization can be calculated perturbatively as
\begin{eqnarray}
V&\sim& h \int dz\cos\left[{f_0\over 2l}(z^2-a^2)\right]\nonumber\\
               &\sim &\sqrt{\xi}\cos\left[{f_0\over 2l}a^2-\frac{\pi}{4}\right].
\end{eqnarray}
Putting in the $\lambda$-dependence and recalling the definition of the conversion measure in Eq.~(\ref{CM2}), we can rewrite the above equation as 
\begin{equation}
    V(\lambda^2)=\sqrt{2}{\rm CM}~\lambda^2\cos\left[{\rm RM}~\lambda^2-{\pi\over 4}\right],
\end{equation}
where $RM$ is the rotation measure determined at the point of reversal. For several reversals the equation above generalizes to
a sum:
\begin{equation}
    V(\lambda^2)=\sqrt{2}\lambda^2~\Sigma_i~ {\rm CM}_i\cos\left[{\rm RM}_i~\lambda^2+\phi _i\right].
    \label{important}
\end{equation}
Here ${\rm RM}_i$ is the rotation measure measured at the $i$'th field reversal, the phase $\phi_i$ is determined by the orientation of the perpendicular component of the magnetic field at the reversal. The equation above is valid in the limit of small conversion, 
and in our opinion is the most useful from a practical point of view. By fitting the observed $V(\lambda^2)$ directly, or by taking a Fourier transform and analyzing the peaks, one can infer the information about the $\rm RM$ and $\rm CM$ of each of the reversal points. This is illustrated in Fig.~\ref{fig:smoo}.

So far we have assumed that the reversals are smooth; however if the field has small-scale structure, it is not a-priori clear that the
peaks in the Fourier transform of $V(\lambda^2)$ survive. This issue is studied in the next subsection.
%The circular polarization Stokes parameter oscillates periodically in $\lambda^2$. The amplitude of the oscillation is $\propto \lambda^2$, as claimed in Eq.~(\ref{rmsp}). 

%It is now clear that if there are several zeros of $B_z$ along the line of site, the observed circular polarization $V$ oscillates quasiperiodically in $\lambda ^2$. The number of different quasiperiods is equal to the number of zeros. The rms degree of circular polarization is still $\propto \lambda^2$, when averaged over a large enough frequency interval. 

\subsection{Turbulent magnetic field}
\label{sec:tur}
If the magnetic field is intrinsic to the plasma, it is most likely turbulent. We will assume the Kolmogorov spectrum 
\begin{equation}\label{kol}
B_r\sim B \left(\frac{r}{l_c}\right)^{1/3}.
\end{equation}
Here $B_r$ is the characteristic random component of the magnetic field at length scale $\sim r$, $l_c$ is the macroscopic length scale that contains most of the magnetic energy, and  $B$ is the characteristic magnetic field at the scale of $l_c$. For simplicity, we take $l_c\sim  l$, where $l$ is the screen thickness.

It would seem that the screen with a turbulent magnetic field is not described by the analysis of \S~\ref{sec:con}. The magnetic field gradient $\sim B_r/r\propto r^{-2/3}$ is now dominated by the magnetic field fluctuations at small scales. Each large-scale zero of $B_z$ splits into infinitely many zeros with an infinite derivative, and the adiabatic invariant is not conserved. The small-scale cutoff of the turbulence might show up, complicating the picture.

\begin{figure}
 \includegraphics[width=0.5\textwidth]{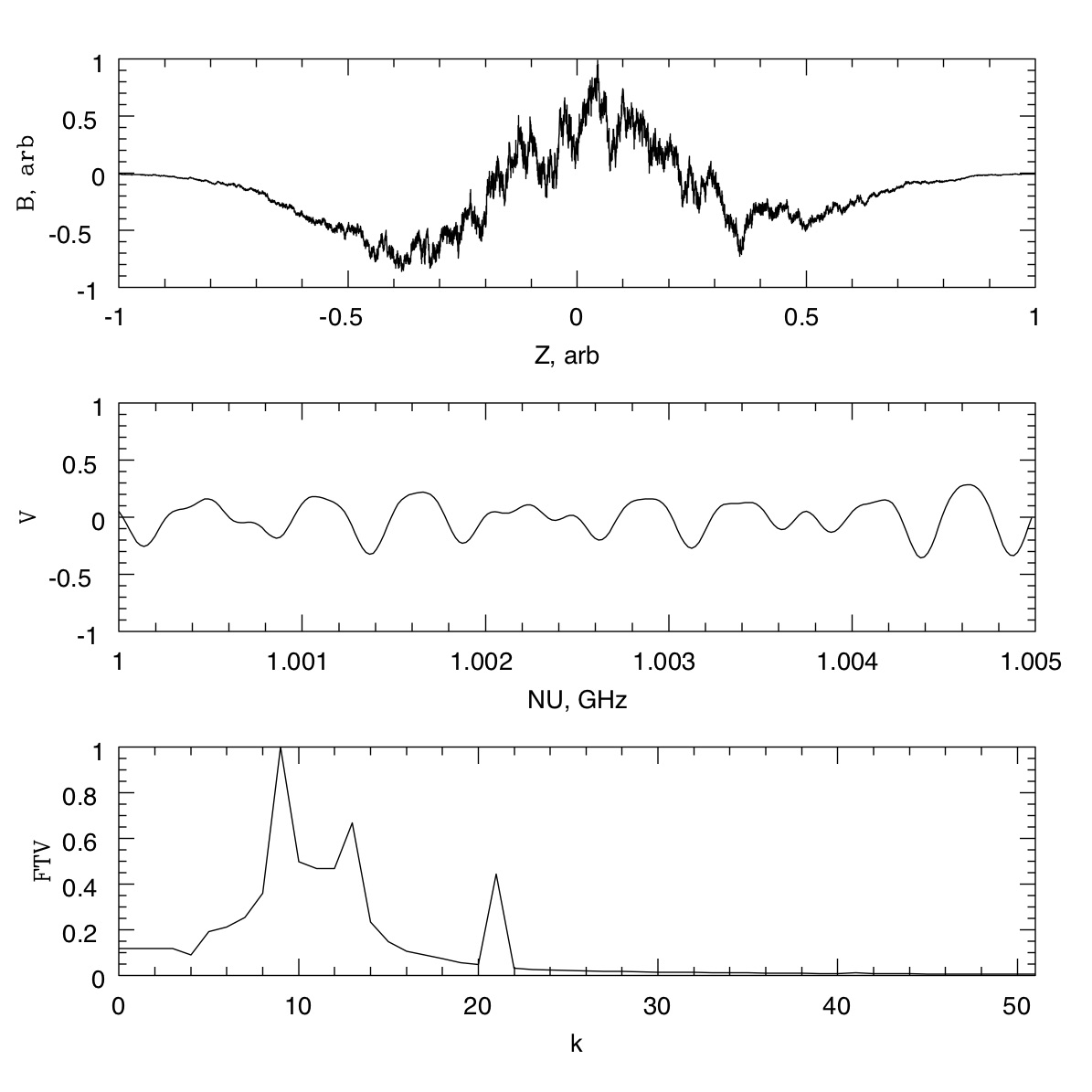}
\caption{Same as Fig.~\ref{fig:smoo}, with Kolmogorov noise added to $B_z$. The peaks are somewhat disturbed, but survive the turbulence.}
\label{fig:turb}
\end{figure}

We numerically integrated the polarization transfer equation (\ref{tra}) for dozens of realizations of a turbulent magnetic field with Kolmogorov spectrum (\ref{kol}). Much to our surprise, we found that the concept of conversion measure survives, as well as the possibility to count the number of ``pronounced zeros'' of $B_z$ by counting the number of quasiperiods of $V$ as a function of $\lambda^2$. An example of such numerical integration is shown in Fig.~\ref{fig:turb}.  We are able to explain this result analytically, as follows:

For ${\nu_B}/{\nu}\ll 1$, we solve Eq.(\ref{tra}) perturbatively:
\begin{equation}
Q+iU=e^{i\alpha \int dz~nB_z},~~~\alpha\equiv \frac{e}{mc^2}\frac{\omega_p^2}{\omega^2},
\end{equation}
\begin{eqnarray}
V&=&\beta {\rm Re} \int dz~n(B_x+iB_y)^2(Q+iU),\nonumber\\
\beta&\equiv& \frac{e^2}{2m^2c^3}\frac{\omega_p^2}{\omega^3}.
\end{eqnarray}
We then calculate the expectation value of $V^2$, assuming Gaussian isotropic parity-invariant magnetic field in the screen
\begin{equation}
\langle B_i({\bf k})B_j({\bf k}')\rangle=(2\pi)^3\delta({\bf k}+{\bf k}')M(k)(\delta_{ij}-\hat{k}_i\hat{k}_j).
\end{equation}
One can show that the magnetic field components along a given line of site, say along $x=y=0$, are independent Gaussian random fields. This gives 
\begin{equation}
\langle V^2 \rangle \sim \beta^2n^2B^4l~\delta l,
\end{equation}
where $\delta l$ is the decorrelation length of $Q+iU$. From 
\begin{equation}
\begin{split}
\langle (Q-iU)|_{z_1}(Q+iU)|_{z_2}\rangle = ~~~~~~~~~~~~~~~~\\ 
 \langle\exp i\alpha \int_{z_1}^{z_2}dz~nB_z\rangle\sim e^{-\frac{1}{2}\alpha^2n^2B^2|z_2-z_1|^2}
\end{split}
\end{equation}
we estimate
\begin{equation}
\delta l\sim \frac{1}{\alpha nB},
\end{equation}
and 
\begin{equation}
V\sim \frac{\nu_B}{\nu}({\rm RM}\lambda ^2)^{1/2}\propto \lambda^2,
\end{equation}
exactly as for the smooth field.

We also see numerically that $V$ as a function of $\lambda^2$ oscillates quasiperiodically, although now the quasiperiodicity is to be understood not as a finite number of incommensurate frequencies as in Eq.~(\ref{important}),  but as a finite number of pronounced peaks in the Fourier transform of $V$ as a function of $\lambda^2$. This happens because the infinite number of zeros of $B_z$ (neglecting the small-scale cutoff) form well defined clusters, with the number of clusters $\sim {l}/{l_c}$. In Fig~3, we show a numerical example of this effect.

\subsection{Non-perturbative regime ${\rm CM} ~\lambda^2\gtrsim 1$.}
\label{sec:lcm}

Although ${\rm CM}~ \lambda^2\lesssim 1$ seems to be more relevant for FRBs as discussed in \S~\ref{sec:appl}, the non-perturbative regime ${\rm CM} ~\lambda^2 \gtrsim 1$ cannot be excluded a priory. In this regime the cases of smooth and turbulent magnetic field are very different. 

For a turbulent field, we find numerically a full Rotation-Conversion regime with $Q\sim U\sim V$ at all frequencies below ${\rm CM} ~\lambda^2\sim 1$. This would have been a natural expectation (as our perturbative result is $V\sim {\rm CM} ~\lambda^2$ and $V$ cannot be greater than 1) were it not at odds with the smooth magnetic field case which we consider next.

In the smooth field case, one needs to break an adiabatic invariant to convert linear into circular polarization, which occurs when
\begin{equation}
\left|\Omega '\right|\gtrsim \Omega^2,
\end{equation}
or, putting a $B_z$ zero at $z=0$, 
\begin{equation}
\frac{f_0}{l}\gtrsim \left(\frac{f_0}{l}z\right)^2+g^2,
\end{equation}
or
\begin{equation}\label{nad}
1\gtrsim ({\rm RM}~\lambda^2)\frac{z^2}{l^2}+({\rm CM}~\lambda^2)^2.
\end{equation}
In the perturbative regime, ${\rm CM} ~\lambda^2\ll 1$, this formula gives the thickness of the non-adiabatic region $z$. But at ${\rm CM} ~\lambda^2\gtrsim 1$, the inequality (\ref{nad}) simply cannot be satisfied -- the screen is everywhere adiabatic. Since adiabatic invariants are conserved to exponential accuracy, one gets an exponential cutoff of conversion at $\lambda > \lambda _c$, where $\lambda _c$ is model dependent, but to order of magnitude given by ${\rm CM} ~\lambda^2_c\sim 1$.

\section{FRB121102}
\label{sec:appl}

What is the magnetic field inside the medium surrounding this source? The FRB is coincident with a 
radio nebula that generates synchrotron radiation with the luminosity of $\sim 10^{39}$erg/s (Chatterjee et al.~2017). Beloborodov [2017, Eq.~(4) of that paper] used the spectral shape of the observed radiation, to derive the magnetic field of $B\sim 0.1$G. This is consistent with the estimate from a one-zone model of Margalit \& Metzger [2018, their Eq.~(17)], which is broadly based on Beloborodov's scenario for powering the nebula and is designed to produce the observed RM. 

It is unknown whether the Faraday screen is located inside or outside the nebula.   
However, the magnetic field of the Faraday screen is strongly constrained by the RM and the dispersion measure of the pulses.  A dramatic reduction of $10$\% in the RM  occurred after the
initial measurement of the linear polarization [Michilli et al.~2018, see section 2 of Vedantham \& Ravi (2018) for a summary]. Since no measurable simultaneous change in the dispersion measure  occurred ($<1 \hbox{pc}\cdot \hbox{cm}^3$), one is able to derive a very conservative lower limit on the mean 
field,
\begin{equation}
     B  > (0.02/\eta_B)\hbox{G},
\end{equation}
see Eq.~(3) of Vedantham \& Ravi (2018). Here $\eta_B$ is the average value of $B_z/B$ along the line of sight, expected to be considerably
smaller than $1$ especially if the medium has field reversals. We shall thus assume 
\begin{equation}
    B\gtrsim 0.1\hbox{G};
    \end{equation}
    it is reassuring that this estimate is consistent with that of the magnetic field in the radio nebula.

%\begin{itemize}

%\item ${\rm RM}\sim 10^5\frac{\rm rad}{{\rm m}^2}$, ~~~~~$\frac{\Delta {\rm RM}}{\rm RM}\sim 1$ in $\sim$few years; 

%\item $\Delta {\rm DM}\lesssim 1{\rm pc}\cdot {\rm cm}^{-3}$ in $\sim$few years, where ${\rm DM}\equiv \int dz~n$ is the dispersion %measure of the screen.

%\item 100\% linear polarization at $\nu \gtrsim 10$GHz.

%\end{itemize}
We can now estimate the CM expected from the linearly polarized pulses of FRB121102.
Our Eq~(\ref{cme}) gives 
\begin{equation}
{\rm CM}\gtrsim 0.3\frac{1}{{\rm m}^2}.
\end{equation}
This was reported in the introduction  as ${\rm CM}\sim {1}/{{\rm m}^2}$ because this is about the median value that we get numerically, mostly due to the partial cancellation of positive and negative RM regions in the Faraday screens with zeros of $B_z$. 

We must stress that Faraday screens with coherence length of order thickness, $l_c\sim l$, have very large scatter of the resulting CM. In particular, $B_z$ may have no zeros and, as a result, no conversion at all would occur,
\begin{equation}
{\rm CM}\approx 0.
\end{equation}
 In other instances, the RM cancellation inside the screen is strong, and one gets 
\begin{equation}
{\rm CM}\sim 10\frac{1}{{\rm m}^2},
\end{equation}
in which case Faraday Rotation at $\nu \sim 1$GHz becomes a non-perturbative Faraday Rotation-Conversion, with observed $Q\sim U\sim V$. 

\section{Discussion}
\label{sec:disc}

This paper provides qualitative predictions for the circular polarization of FRB pulses, if it is produced by Faraday Conversion of an initially linearly polarized pulses, such as the ones in FRB121102. If the FC occurs in cold plasma, then it takes place near the field reversals along the line of sight. Each field reversal produces a quasi-period in $V(\lambda^2)$, with the frequency and amplitude containing information about the Rotation Measure and Conversion Measure of the reversal, respectively. Our work motivates narrow-band full polarization measurements at low frequencies; the pay-off is the measurement of the architecture of the magnetic environment of the 
FRB, as well as a confirmation of the beautiful physics of Faraday conversion in cold plasmas.

The discussion of this paper is incomplete; we have concentrated only on cold plasma. The FC in relativistic plasma with non-trivial magnetic geometry will be addressed in future work.
%An interesting regime of Faraday Rotation-Conversion may show up in FRBs, especially in the repeating FRB121102 at low frequencies. It may be worth observers' efforts (very narrow bandwidths) to measure the low-frequency polarization, because in the Rotation-Conversion regime the Faraday screen does not reduce to a single number RM, but contains additional information which might help understand FRBs.

\vskip 1cm

\acknowledgements 

AG thanks the many participants of 2018 Weizmann FRB workshop for useful information. {Both authors thank Harish Vedantham for initiating their interest in Faraday Conversion, for detailed comments on the manuscript, and for pointing out that some of the results have been previously obtained by Zheleznyakov \& Zlotnik (1964).} YL thanks Andrei Beloborodov for useful discussions.

\vskip 1cm

\appendix

\section{Rotation-Conversion in a cold magnetized plasma}

Rotation-Conversion in a cold magnetized plasma is simple because there is no emission and absorption. To calculate the effect one proceeds along standard lines -- calculate the permittivity and then the radiation transfer. A common reference is \cite{Gin}. We do it below in full and in somewhat different terms. 

\subsection{Permittivity}
Let ${\bf B}$ be the background constant uniform magnetic field, ${\bf E}$ is the electric field of the wave, $\omega$ is the frequency of the wave. A cold electron moves according to Lorentz equation ($c=1$ here and below)
\begin{equation}
m\dot{\bf v}=e({\bf E}+{\bf v}\times {\bf B}),
\end{equation}
or
\begin{equation}
-i\omega {\bf v}=\frac{e}{m}({\bf E}+{\bf v}\times {\bf B}),
\end{equation}
or
\begin{equation}
{\bf v}=i\frac{e}{m}\frac{\omega}{\omega^2-\omega_B^2}\left( {\bf E}-i\frac{\omega_B}{\omega}\hat{B}\times {\bf E}-\frac{\omega_B^2}{\omega^2}\hat{B}(\hat{B}\cdot {\bf E})\right),~~~~~\omega_B\equiv \frac{eB}{m},~~~~~\hat{B}\equiv\frac{\bf B}{B}.
\end{equation}

Moving electrons create plasma current ${\bf j}=ne{\bf v}$, ions contribute much less and are neglected. The dielectric permittivity tensor $\epsilon$ is, by definition, given by $4\pi {\bf j}+\partial_t{\bf E}\equiv \partial_t(\epsilon {\bf E})$, or $-i\omega (\epsilon -1){\bf E}=4\pi {\bf j}$, and we get
\begin{equation}
\epsilon _{ij}=\epsilon_\perp\delta_{ij}+(\epsilon_\parallel-\epsilon_\perp)\hat{B}_i\hat{B}_j+ige_{ijk}\hat{B}_k.
\end{equation}
Here $e_{ijk}$ is the antisymmetric unit tensor and 
\begin{equation}
\epsilon_\perp=1-\frac{\omega_p^2}{\omega^2-\omega_B^2},~~~~~\epsilon_\parallel=1-\frac{\omega_p^2}{\omega^2},~~~~~g=-\frac{\omega_p^2\omega_B}{(\omega^2-\omega_B^2)\omega}, ~~~~~\omega_p^2\equiv \frac{4\pi ne^2}{m}.
\end{equation}
As $\omega\gg \omega_B, \omega_p$ is assumed, we can replace, to sufficient accuracy,
\begin{equation}
\epsilon_\perp\approx 1-\frac{\omega_p^2}{\omega^2}+\frac{\omega_p^2\omega_B^2}{\omega^4},~~~~~g\approx-\frac{\omega_p^2\omega_B}{\omega^3}. 
\end{equation}

\subsection{Radiation Transfer}
Maxwell equations give the eigenmode equation
\begin{equation}
(k^2\delta_{ij}-k_ik_j-\omega^2\epsilon_{ij})E_j=0,
\end{equation}
or
\begin{equation}\label{eig}
(k^2-\omega^2\epsilon_\perp)E_i=k_ik_jE_j+\omega^2\left( ige_{ijk}\hat{B}_k+(\epsilon_\parallel-\epsilon_\perp)\hat{B}_i\hat{B}_j\right)E_j, 
\end{equation}
where ${\bf k}$ is the wavevector. Consider a wave propagating along $z$: ${\bf k}=(0,0,k)$. To sufficient accuracy, one can neglect $E_z$ in the first two equations of Eq.(\ref{eig}), and also use $\epsilon_\perp\approx 1-\frac{\omega_p^2}{\omega^2}$ in the l.h.s. of Eq.(\ref{eig}). Then
\begin{equation}
(k^2+\omega_p^2-\omega^2)E_a=-\omega^2\left( i\frac{\omega_p^2\omega_B}{\omega^3}\hat{B}_ze_{ab}+\frac{\omega_p^2\omega_B^2}{\omega^4}\hat{B}_a\hat{B}_b\right)E_b,
\end{equation}
where the indices $a$, $b$ run from 1 to 2, and $e_{ab}$ is the 2D antisymmetric unit tensor. 

Replacement
\begin{equation}
k=\sqrt{\omega^2-\omega_p^2}-i\partial_z
\end{equation}
gives the polarization transfer equation for complex amplitudes $E_a$:
\begin{equation}\label{camp}
\partial_zE_a=\frac{1}{2}\left(\frac{\omega_p^2\omega_B}{\omega^2}\hat{B}_ze_{ab}-i\frac{\omega_p^2\omega_B^2}{\omega^3}\hat{B}_a\hat{B}_b\right)E_b.
\end{equation}

Defining Stokes parameters,
\begin{equation}
E_aE_b^*\equiv\frac{1}{2}\left(
\begin{array}{cc}
I+Q & U+iV\\

\\

U-V & I-Q\\

\end{array}
\right),
\end{equation}
and using Eq.(\ref{camp}) we get Eq.(\ref{tra}).

\vskip 1cm

\bibliographystyle{hapj}

\end{document}